\documentclass[conference]{IEEEtran}

\usepackage{cite}
\usepackage{amsmath,amssymb,amsfonts}
\usepackage{algorithmic}
\usepackage{graphicx}
\usepackage{textcomp}
\usepackage{xcolor, soul}

\usepackage{float}
\usepackage{tabularx}
\usepackage{flushend}

\def\BibTeX{{\rm B\kern-.05em{\sc i\kern-.025em b}\kern-.08em
    T\kern-.1667em\lower.7ex\hbox{E}\kern-.125emX}}

\begin{document}

\title{Circuit Theory-Informed AI for Condition Monitoring in Power Electronics}

\author{
    \IEEEauthorblockN{Yubo~Song\textsuperscript{1}, Pascal~A.~Schirmer\textsuperscript{2}, and Subham~Sahoo\textsuperscript{1}}
    \IEEEauthorblockA{
        \textsuperscript{1}\textit{Department of Energy, Aalborg University, Aalborg, Denmark} \\
        \textsuperscript{2}\textit{Research and Innovation Center, BMW Group, Munich, Germany} \\
        \textsuperscript{1}\{\textit{yuboso, sssa}\}@\textit{energy.aau.dk}, \textsuperscript{2}\textit{pascal.schirmer}@\textit{bmw.de}
    }
}

\maketitle

\begin{abstract}
Artificial intelligence for power electronics is evolving toward a deeper comprehension of system intrinsic behaviors and dynamics. As a novel integration of neural computation and electric circuits, spike-based neuromorphic computing offers promising potential in high energy efficiency and adaptability for data-driven applications. Building on the latency-coding principles, this article proposes a framework for data-light condition monitoring that relies solely on three-phase AC current measurements. The AC current signals are encoded as membrane potentials of RC-based spiking neurons, which are inherently compatible with spiking neural networks that can effectively facilitate the estimation of device parameters. Case studies are presented to validate the performances, while constraints and further optimizations are also evaluated and discussed, and future extensions of the methodology are provided.
\end{abstract}


\section{Introduction}

The digitalization of power electronics systems is transforming not only how information should be processed but also how such systems should be perceived. This goes beyond purely hardware-based approaches towards the comprehension of the intrinsic behaviors of passive circuits in power electronic systems. In this context, artificial intelligence (AI) has become a powerful enabler for data-driven analysis in power electronics, owing to its ability to handle complex nonlinearities \cite{Minjie2025AI, Shuai2021AI}. AI-based methods are increasingly unveiling subtle patterns in power electronics systems when they are inaccessible by conventional model-based approaches.

\begin{figure}[t]
    \centering
    \includegraphics[width=0.8\linewidth]{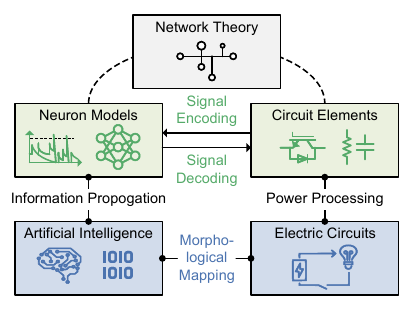}
    \vspace{-6pt}
    \caption{Network theory as the bridge between neural networks and electric circuits. Signals are mapped via encoding/decoding algorithms, and power flow serves as an interpretation of neural signal propagation.}
    \label{fig_circuitai}
\end{figure}

This trend raises a research question on how to exploit the shared physics between AI and electric circuits, as in Fig.~\ref{fig_circuitai}. Signal propagation in neural networks and interconnections in analog circuits are both subject to the principles of network theory, while signal encoding and decoding offer morphological duality and the potential for more efficient data processing by leveraging circuit-level dynamics, for instance, networked memristors as in-memory computing solutions \cite{Osvaldo2020Memristor}. Similarly, this analogy is also promising for power electronics systems at a macro scale with high controllability.

The inspiration from biological neural systems to neuromorphic computing \cite{Yu2017Neuromorphic, Elishai2022NeuroEng, Nature2022Neuromorphic} can be leveraged in this context, where, unlike digital computation, information is transmitted as asynchronous impulses (known as spikes) in spiking neural networks (SNNs) rather than uniformly-sampled floating-point values. This \textit{event-driven} mechanism enables energy-efficient information processing with sparse data \cite{LowPowerMag, TrueNorth}, as well as high adaptability through inter-neuron dependencies \cite{TrainingSNN}. Meanwhile, various neuromorphic spike-encoding schemes, such as rate, latency and burst coding \cite{TrainingSNN}, have not only linked SNNs to general data-processing tasks, but also provide an alternative means of learning the system dynamics \cite{SpikeTalk, SpikeTalkMag}.

For power electronics, condition monitoring remains challenging in terms of high requirements on advanced hardware sensors \cite{Forest2013, Peng2021Von, Chen2024} or high-fidelity data for deep learning or digital twin approaches \cite{Shuai2020DataDrivenCM, Shuai2022DT}. For instance, monitoring the \textsc{on}-state voltage drop $\mathrm{V_{CE}}$ of power semiconductor devices as an indicator of the increasing \textsc{on}-state channel resistance $\mathrm{R_{dson}}$ during device degradation \cite{Oh2015PoF}. Although these methods can achieve satisfactory modeling accuracy, their applicability in industrial settings can be limited by energy consumption, reliability and data availability \cite{Shuai2020DataDrivenCM}.

This article thereby explores neuromorphic concepts for condition monitoring in power electronics, focusing on latency coding to alleviate data-sensing requirements in conventional approaches. Relying solely on AC current measurements at an adequate frequency, the proposed framework leverages spiking neurons to effectively encode current signals into spike patterns that are highly compatible with energy-efficient SNNs, thereby facilitating the estimation of $\mathrm{R_{dson}}$ in power electronics converters. Rather than positioning spiking layers as an efficient accelerator alongside physical fitting \cite{baik2026}, we aim to assess the capability of SNNs as the independent core. Representative case studies are presented, and results are given to demonstrate the principles and effectiveness of the methodology and evaluate its capability. Ultimately, in-depth discussions are provided outlining the future extensions of the methodology and the intrinsic physics.

\section{Neuromorphic Modeling Principles}

\begin{figure}[t]
    \centering
    \includegraphics[width=\linewidth]{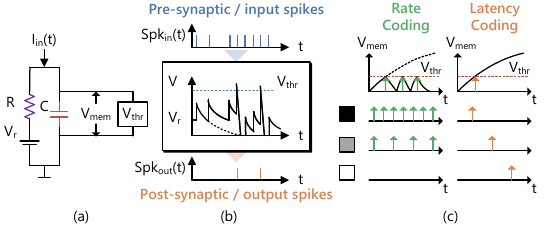}
    \caption{Essential physics of neuromorphic computing \cite{TrainingSNN}. (a) Typical spiking neuron model with $\mathrm{RC}$ topology and the spike-firing logic. (b) Rate coding and latency coding, where signals are encoded as firing rate and time-to-first-spike of the spikes, respectively. Larger inputs correspond to higher rates or earlier occurrences of the first spike.}
    \label{fig_neuron}
\end{figure}

Neuromorphic computing primarily originates from the physics of biological neurons \cite{Elishai2022NeuroEng}. Information is transferred through synapses, and the post-synaptic neuron is activated when the membrane potential reaches the threshold due to sufficient stimuli from the pre-synaptic neurons. In computational neuroscience, this paradigm can be modeled as an $\mathrm{RC}$ circuit (Fig.~\ref{fig_neuron}(a)), the leaky integrate-and-fire (LIF) neuron model, where the capacitor $\mathrm{C}$ is charged by the input current $\mathrm{I_{in}}$ (stimuli), but meanwhile decays naturally due to the resistance $\mathrm{R}$. This dynamics can be written as follows \cite{SpikeTalk}:
\begin{equation}
    \mathrm{I_{in}(t) = \frac{V_\mathrm{mem}(t) - V_r}{R} + C\frac{\mathrm{d}V_\mathrm{mem}}{\mathrm{d}t}}
\end{equation}
where, $\mathrm{V_r}$ is the relaxation voltage (assumed as zero), and $\mathrm{\tau=RC}$ is the time constant.

A spike is triggered when the membrane potential $\mathrm{V_\mathrm{mem}}$ hits the threshold $\mathrm{V_\mathrm{thr}}$ (Fig.~\ref{fig_neuron}(b)):
\begin{equation}
    \mathrm{S_{out}}=\left\{
    \begin{array}{rcl}
    1 & & \mathrm{V_\mathrm{mem}\geq V_\mathrm{thr}}\\
    0 & & \mathrm{V_\mathrm{mem}<V_\mathrm{thr}}
    \end{array} \right.
\end{equation}
and the output spikes are triggered dependent on the accumulation of input spikes (Fig.~\ref{fig_neuron}(b)), accounting for both the $\mathrm{RC}$ decay and the threshold $\mathrm{V_\mathrm{thr}}$.

SNN is formulated upon this neuron model as the tool for neuromorphic computing, where the data are encoded into spikes trains (Fig.~\ref{fig_neuron}(c)) and then transmitted through neurons. Meanwhile, as it follows an \textit{event-driven} paradigm, it exhibits remarkable potentials in energy efficiency ane enables unique training strategies distinct from conventional artificial neural networks \cite{TrainingSNN}.

\section{Neuromorphic Inference for Condition Monitoring}

\subsection{Spiking Encoding of AC Signal}

\begin{figure}[t]
    \centering
    \includegraphics[width=\linewidth]{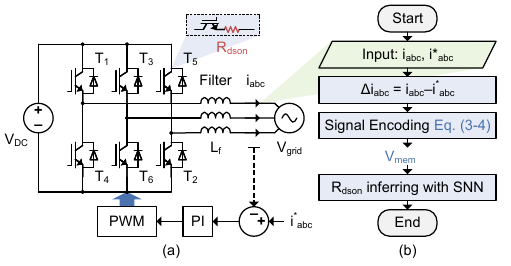}
    \caption{Study case of this article. (a) A grid-connected DC-AC converter is exemplified. (b) Framework of $\mathrm{R_{dson}}$ inference for the study case.}
    \label{fig_circuit}
\end{figure}

Following the principles of spiking neuron modeling, we subsequently propose a circuit theory-informed approach for condition monitoring. In this article, we focus on the inference of the \textsc{on}-state resistances $\mathrm{R_{dson}}$ of power semiconductors (which increases during device degradation \cite{Oh2015PoF}) by exemplifying a grid-connected DC-AC converter in Fig.~\ref{fig_circuit}(a) through PLECS simulations, with key parameters specified in Table~\ref{tab_par}. As a numerical demonstration of the methodology, we define $\mathrm{R_{dson}}$ as the increase from ideal initial state where $\mathrm{R_{dson}}$ of all devices equal to zero, and the $\mathrm{R_{dson}}$ of each device is analyzed up to 1~$\Omega$, without loss of generality. The overall inference framework is outlined in Fig.~\ref{fig_circuit}(b).

\begin{table}[t]
    \renewcommand{\arraystretch}{1.25}
    \centering
    \footnotesize
    \caption{Rated Parameters of the Studied System}
    \label{tab_par}
    \vspace{-6pt}
    \begin{tabular}{p{0.35\linewidth} >{\centering\arraybackslash}p{0.15\linewidth}}
    \hline\hline
    \multicolumn{1}{c}{Parameters} & Values \\
    \hline
    Grid Voltage (rms) $\mathrm{V_\mathrm{g}}$ & 110~V \\
    DC Voltage $\mathrm{V_\mathrm{DC}}$ & 400~V \\
    Fundamental Frequency $\mathrm{f_\mathrm{1}}$ & 50~Hz \\
    Switching Frequency $\mathrm{f_\mathrm{sw}}$ & 10~kHz \\
    Current Rating (rms) $\mathrm{I_\mathrm{abc}^*}$ & 10~A \\
    Filter Inductance $\mathrm{L_\mathrm{f}}$ & 5~mH \\
    \hline\hline
    \end{tabular}
\end{table}

The core of the framework lies in detecting system fluctuations around the rated operation state. Given that the closed-loop control that regulates the AC currents $\mathrm{\mathbf{i}_{abc}}$, an increase of $\mathrm{R_{dson}}$ will lead to observable changes in current fluctuations relative to rated sinusoidal references $\mathrm{\mathbf{i}_{abc}^*}$. By integrating the mean-square-errors (MSE) of $\mathrm{\mathbf{i}_{abc}}$ in terms of $\mathrm{\mathbf{i}_{abc}^*}$ for each phase, we have the input to an LIF neuron:
\begin{equation}
    \label{eq_Iin}
    \mathrm{\mathbf{I}_{in, abc}(t) = \int_0^t{(\mathbf{i}_{abc}-\mathbf{i}_{abc}^*)^2\mathrm{d}t}}
\end{equation}
It should be noted that (\ref{eq_Iin}) only stands for a numerical mapping disregarding physical units. Then the corresponding membrane potential $\mathrm{\mathbf{V}_{mem}}$ is obtained by passing $\mathrm{\mathbf{I}_{in}}$ through the $\mathrm{RC}$ low-pass filter (LPF):
\begin{equation}
    \label{eq_Vmem}
    \mathrm{\mathbf{V}_{mem} = \mathbf{I}_{in}\cdot\frac{R_{enc}}{1+s\tau_{enc}}}
\end{equation}

We further employ the latency (temporal) coding scheme in neuromorphic computing \cite{TrainingSNN}, in which the magnitude of a signal is encoded as the time-to-first-spike within a spike train. For an $\mathrm{RC}$ neuron model as illustrated in Fig.~\ref{fig_neuron}(a) and given step $\mathrm{I_{in}}$ as the charging current, the time-to-first-spike follows first-order $\mathrm{RC}$ dynamics and can be derived as \cite{SpikeTalk}:
\begin{equation}
    \label{eq_latency}
    \mathrm{t(I_{in})} =
    \begin{cases}
        \mathrm{\tau\left[\ln{\left(\cfrac{I_{in}R}{I_{in}R-V_\mathrm{thr}}\right)}\right]},\quad \textbf{if}\; \mathrm{I_{in}R > V_\mathrm{thr}} \\[3pt]
        \infty,\quad \textbf{else}
    \end{cases}
\end{equation}
where higher input currents yield earlier spike occurrences. Further considering the sensitivity of $\mathrm{t(I_{in})}$ with respect to $\mathrm{I_{in}}$, (\ref{eq_sensitivity}) can be obtained:
\begin{equation}
    \label{eq_sensitivity}
    \mathrm{\frac{\partial t(I_{in})}{\partial I_{in}} = -\frac{\tau V_\mathrm{thr}}{I_{in}(I_{in}R-V_\mathrm{thr})}}
\end{equation}
Although (\ref{eq_latency}) is derived under a constant $\mathrm{I_{in}}$, it also reveals the contribution of tunable parameters $\mathrm{R}$, $\mathrm{\tau=RC}$ and $\mathrm{V_\mathrm{thr}}$ to the sensitivity of $\mathrm{R_{dson}}$ detection (via $\mathrm{I_{in}}$) and subsequent optimization of SNN training.

\begin{figure}[t]
    \centering
    \includegraphics[width=\linewidth]{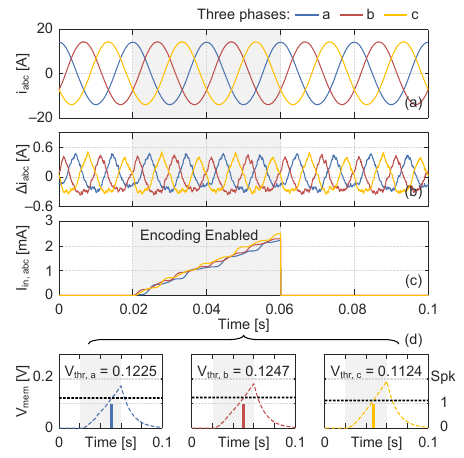}
    \vspace{-12pt}
    \caption{Illustration of the spike encoding approach when $\mathrm{R_{dson1} = 0.2\:\Omega}$. (a) Three-phase sinusoidal current, (b) three-phase current fluctuation, (c) the input current for the encoding neuron as calculated by (\ref{eq_Iin}), and (d) membrane potentials (dashed lines) and latency-encoded spikes (solid lines) where the thresholds $\mathrm{\mathbf{V}_{thr, abc}}$ are set such that the spikes are aligned among phases when all $\mathrm{R_{dson}= 0\:\Omega}$. 
    The $\mathrm{RC}$ neuron is specified as: $\mathrm{R=100\:\Omega}$ and $\mathrm{\tau=10\:ms}$.
    }
    \label{fig_current}
\end{figure}

An example of the encoding process is illustrated in Fig.~\ref{fig_current}, where $\mathrm{\Delta\mathbf{i}_\mathrm{abc}}$ represents the fluctuation of three-phase AC currents $\mathrm{\mathbf{i}_\mathrm{abc}}$ from its reference values $\mathrm{\mathbf{i}_\mathrm{abc}^*}$, which is used for constructing $\mathrm{\mathbf{I}_{in}}$ as per (\ref{eq_Iin}) and to
obtain $\mathrm{V_{mem}}$ through the $\mathrm{RC}$ neuron model (Fig.~\ref{fig_current}(c) and (d)). In this example, the encoding is performed over 2 fundamental periods (from 0.02 to 0.06 s) at a sampling frequency of $\mathrm{f_s=f_{sw}=10\:kHz}$. Without loss of generality, the thresholds $\mathrm{\mathbf{V}_{thr, abc}}$ are set such that the spikes are aligned among phases when $\forall \mathrm{R_{dson}= 0\:\Omega}$, the resulting encoded spikes are depicted as Fig.~\ref{fig_current}(d).

\begin{figure}[t]
    \centering
    \includegraphics[width=\linewidth]{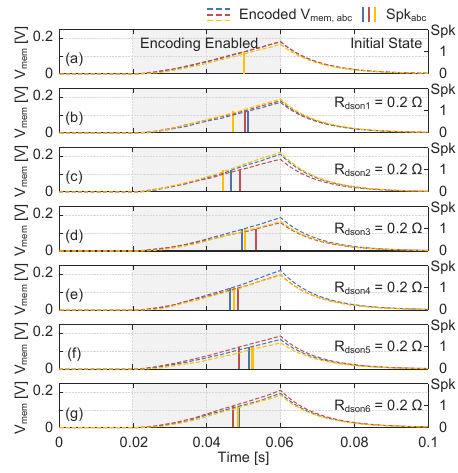}
    \vspace{-24pt}
    \caption{Exemplary encoded membrane potentials $\mathrm{V_{mem}}$ and spikes $\mathrm{Spk}$ of (a) the initial state, and (b)-(g) cases where one specific $\mathrm{R_{dson}}$ (of different devices in each case) increases compared to the initial state. A spiking neural network (SNN) is thereby employed to learn from the spikes.}
    \label{fig_spkalign}
    \vspace{-6pt}
\end{figure}

The effectiveness of this encoding approach is further demonstrated in Fig.~\ref{fig_spkalign}. Test cases in Fig.~\ref{fig_spkalign}(b)-(g) with $\mathrm{R_{dson}}$ increments in different devices but identical encoding configurations indicate that the resulting spike alignments can be directly utilized for SNN-based computations, which is energy-efficient for model inference. Moreover, since the $\mathrm{R_{dson}}$ profiles can be distinguished with only AC signal sampling, the proposed approach also exhibits favorable advantages in low data requirements and sensor dependency.

\subsection{Neuromorphic Inference of ${{R}_{dson}}$}

\begin{figure}[t]
    \centering
    \includegraphics[width=0.85\linewidth]{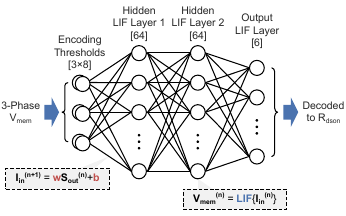}
    \vspace{-6pt}
    \caption{Illustration of the layered SNN applied in this article. To enhance the training performances, multiple thresholds (8 in this article) are employed for each phase and generating spike trains each in length of 100. The sequences from the $3\times 8$ neurons in the encoding layer are then processed by the SNN throughout the following LIF layers.}
    \label{fig_snn}
\end{figure}

\begin{figure}[t]
    \centering
    \includegraphics[width=\linewidth]{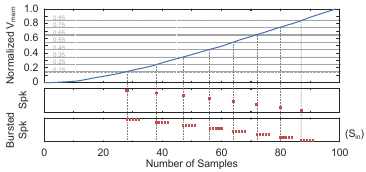}
    \vspace{-9pt}
    \caption{Input spike trains are normalized and encoded from respective thresholds as per the latency-based scheme, and then expanded into bursts to mitigate the sparsity and enhance their representational capacity.}
    \label{fig_spkin}
\end{figure}

The SNN is constructed based on the spike response model (SRM) in \cite{Osvaldo2021SNN}, as presented in Fig.~\ref{fig_snn}. The SNN infers $\mathrm{R_{dson}}$ of all devices from the spike trains encoded by the input current $\mathrm{I_{in}}$. The input current $\mathrm{I_{in}}$ of each phase is encoded using a latency-based scheme in (\ref{eq_Iin}) with multiple thresholds (from 0.15 to 0.85 in our case, 8 in total) in parallel, and the time-to-first-spike representation is then normalized to spike trains, each with sequence length of 100 as the output of a single neuron in the encoding layer in Fig.~\ref{fig_snn}. Alternatively, for sparse inputs, spikes can be grouped into bursts of multiple rather than encoded as single, thereby enhancing the representational capacity, which, in this paper, is applied to the input spike sequences at a fixed burst length of 5, as illustrated in Fig.~\ref{fig_spkin}. After encoding, the entire neural network consists of 3 layers with LIF neurons (LIF layers) and trainable linear scaling applied to the spike transmissions between adjacent layers, processing the spike sequences from the encoding layer throughout the LIF layers. The predicted $\mathrm{R_{dson}}$ of all devices are decoded from the output LIF layer, by combining the membrane potentials as well as the spiking rates, eventually for loss evaluation.

\begin{table}[t]
    \renewcommand{\arraystretch}{1.10}
    \centering
    \footnotesize
    \caption{Specifications of the Spiking Neural Network}
    \label{tab_snn}
    \vspace{-6pt}
    \begin{tabular}{@{\hspace{1em}}p{0.3\linewidth} @{\hspace{1.5em}}>{\centering\arraybackslash}p{0.32\linewidth}@{\hspace{1em}}}
    \hline\hline
    \multicolumn{1}{c}{\textbf{Specifications}} & \textbf{Values} \\
    \hline
    Encoded Time Window & $\mathrm{40\:ms}$ \\
    Encoded Seq. Length & $\mathrm{100}$ \\
    \hline
    LIF Decay Rate & $\mathrm{\beta=0.95}$ \\
    Surrogate Gradient & $\mathrm{Fast\ Sigmoid}$ \\
    & $(\mathrm{slope = 25})$ \\
    \hline
    Training Batch Size & $\mathrm{32}$ for basic cases \\[-1pt]
    & $\mathrm{256}$ for extended cases\textsuperscript{*} \\[1pt]
    Training Optimizer & $\mathrm{AdamW}$ \\
    Initial Learning Rate & $\mathrm{lr=5e\!-\!4}$ \\
    Weight Decay Coeff. & $\mathrm{1e\!-\!4}$ \\
    \hline\hline
    \multicolumn{2}{@{\hspace{1em}}l}{*Note: Validations and evaluations in Section IV.}
    \end{tabular}
    \vspace{-12pt}
\end{table}

\begin{figure}[t]
    \centering
    \includegraphics[width=0.9\linewidth]{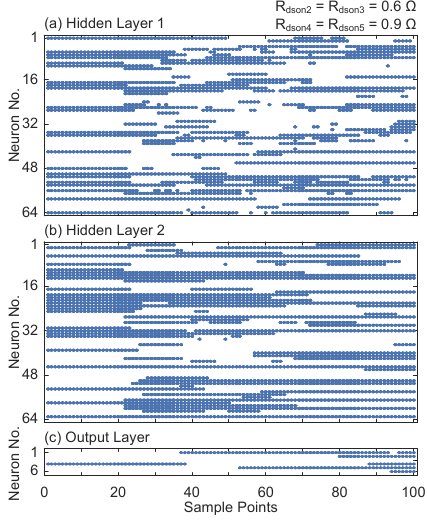}
    \vspace{-6pt}
    \caption{Exemplary spikes at each layer of the SNN trained under the specified conditions. An intermediate spiking rate should be ensured throughout the training, so as to avoid excessive noisy or inactive neurons for both effective and stable outputs.}
    \label{fig_spklyrs}
\end{figure}

The SNN is basically trained with a uniformly sampled dataset ($\mathrm{\mathbf{R}_{dson}} = [0,\:0.3,\:0.6,\:0.9]\:(\Omega)$ for each device, yielding a total of 4096 training samples) for up to 400 epochs. For specified extended validations and evaluations in Section III and IV, larger dataset is adopted yet with increased batch size: The dataset is further expanded with 4096 more samples with random $\mathrm{\mathbf{R}_{dson}}$, and the dataset is shuffled and used for both training (80\% of the samples) and testing. All data are collected through PLECS simulation. Key training parameters are specified in Table~\ref{tab_snn}: \textit{Fast Sigmoid} is used as the surrogate gradient of the SNN, and the learning rate is variable over training epochs to ensure adequate model convergence. Besides, to ensure effective and stable outputs of the SNN, spike activities are observed, with a example visualized in Fig.~\ref{fig_spklyrs}, where spike bursting and regularization contribute to achieving intermediate firing rates generally without excessive noisy or inactive neurons.

\subsection{Basic Performance Validations}

\begin{figure}[t]
    \centering
    \includegraphics[width=0.9\linewidth]{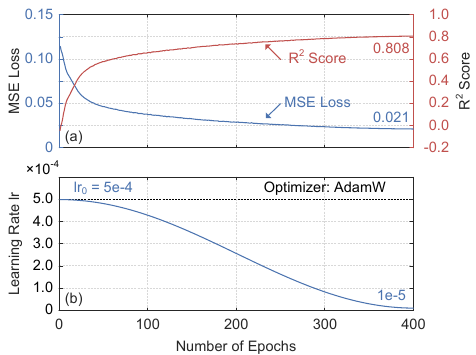}
    \vspace{-6pt}
    \caption{Training metrics of the SNN in relation to the number of epochs: (a) mean-square-error (MSE) loss and $\mathrm{R^2}$ score, and (b) learning rate. Training metrics at 200 epochs are listed in Table~\ref{tab_score}.}
    \label{fig_nnloss}
    \vspace{-6pt}
\end{figure}

The training of the approach are first evaluated in Fig.~\ref{fig_nnloss} and Table~\ref{tab_score}, where the MSE loss gradually decreases when trained for more epochs, meanwhile the $\mathrm{R^2}$ score reaches over 0.8 that confirms the correlations between the encoded inputs and output predictions.

\begin{figure}[t]
    \centering
    \includegraphics[width=0.8\linewidth]{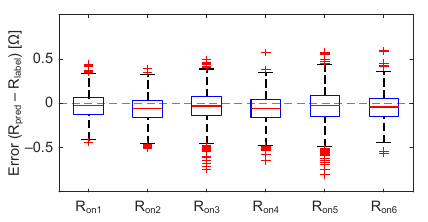}
    \vspace{-3pt}
    \caption{Testing of prediction accuracy with 1000 random samples after training for 400 epochs. The 25\textsuperscript{th} and 75\textsuperscript{th} percentiles are outlined.}
    \label{fig_pred}
\end{figure}

\begin{figure}[t]
    \centering
    \includegraphics[width=\linewidth]{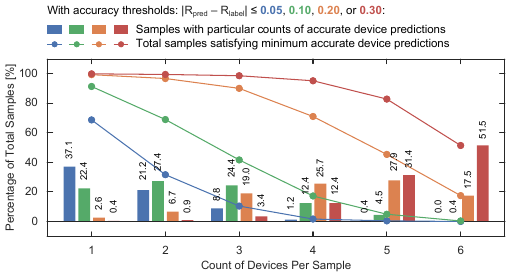}
    \vspace{-12pt}
    \caption{Distribution of prediction accuracy after training for 200 epochs under different defined accuracy tolerance. The validations are performed with the extended dataset (8192 samples in total as specified in Section III-B).}
    \label{fig_acc_dist}
    \vspace{-9pt}
\end{figure}

Further assessment of the prediction accuracy is performed in Fig.~\ref{fig_pred} considering 1000 test samples with arbitrary $\mathrm{R_{dson}}$ (up to 1~$\Omega$) of all devices. The majority of samples align with the pre-set $\mathrm{R_{dson}}$ labels, while the neural network can still benefit from being trained with larger datasets or finer tuning for a learning process with faster and more uniform rate of convergence, eventually towards further enhanced accuracy. The distribution of prediction accuracy is also illustrated in Fig.~\ref{fig_acc_dist}, which statistically reflects the model performance and aligns with Fig.~\ref{fig_pred}.

\begin{table}[!t]
    \renewcommand{\arraystretch}{1.10}
    \centering
    \footnotesize
    \caption{Representative Training Metrics}
    \label{tab_score}
    \vspace{-6pt}
    \begin{tabular}{@{\hspace{1ex}} >{\centering\arraybackslash}p{0.16\linewidth} @{\hspace{1ex}} >{\centering\arraybackslash}p{0.16\linewidth} @{\hspace{1ex}} >{\centering\arraybackslash}p{0.16\linewidth} @{\hspace{1ex}} >{\centering\arraybackslash}p{0.16\linewidth} @{\hspace{1ex}} >{\centering\arraybackslash}p{0.16\linewidth} @{\hspace{1ex}}}
    \hline\hline
    \textbf{Epochs} & \textbf{lr} & \textbf{$\mathrm{R^2}$ Score} & \textbf{MSE Loss} & \textbf{MAE Loss}\\
    \hline
    10 & 4.994$\mathrm{e}$-4 & 0.2306 & 0.0841 & 0.2438\\
    20 & 4.973$\mathrm{e}$-4 & 0.3625 & 0.0697 & 0.2210\\
    50 & 4.821$\mathrm{e}$-4 & 0.5414 & 0.0501 & 0.1823\\
    100 & 4.296$\mathrm{e}$-4 & 0.6388 & 0.0395 & 0.1590\\
    200 & 2.569$\mathrm{e}$-4 & 0.7007 & 0.0327 & 0.1435\\
    \hline\hline
    \end{tabular}
    \vspace{-6pt}
\end{table}

\section{Discussions}

Going beyond the results, we also recognize the following technical considerations worth discussions, as well as possible extensions of the methodology for better model performance and generalization:
\begin{enumerate}
    \item \textit{Data availability}: As a fundamental evaluation of the energy-efficient SNN for long-term condition monitoring applications, the methodology has been deliberately formulated with minimum data requirements, relying solely on AC-side measurements and at low sampling rate. Nevertheless, the accuracy could be improved by utilizing additional signals before encoding such as the DC-link voltage and the PWM switching signals---e.g., as voltage level references or sampling triggers, neither of which adds significantly to the measurement complexity. Such signals would be particularly beneficial for discriminating among switches whose variation produce similar signatures in the said AC signals. They could also enhance the noise resiliency of the model, especially in terms of the common-mode disturbances on parallel measurements.

    Another application is indicated by Fig.~\ref{fig_pred_grp}, which shows the prediction accuracy by device clusters. Geometric symmetry of devices may lead to similar spike patterns and inaccurate device-level predictions, but this issue could be overridden by clustered module-level prediction. Such approach could in turn benefit from the capability of the data-light neuromorphic paradigm and thus improve the data and computing efficiency for condition monitoring.

    \begin{figure}[t]
        \centering
        \includegraphics[width=0.8\linewidth]{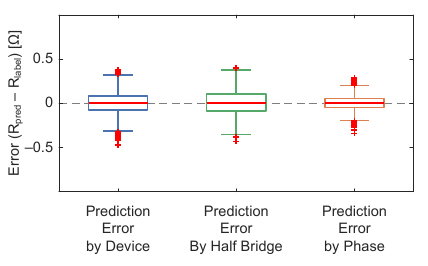}
        \vspace{-3pt}
        \caption{Prediction accuracy by device clusters in comparison with single ones, where the devices are grouped by half bridges (positive/negative halves) or phases (a/b/c phases), and the SNN is trained by summing up device $\mathrm{R_\mathrm{dson}}$ in each cluster. It should be noted that a basic training dataset is exemplified here.}
        \label{fig_pred_grp}
    \end{figure}
    
    \item \textit{Signal encoding methods}: Signal encoding is the second key to improving and generalizing this neuromorphic approach. Here we have converted AC signals into DC membrane potentials through MSE integration to accommodate latency coding, while advanced encoding methods such as \textit{Delta Modulation} \cite{7128412, 10675361} could also be applied (particularly to alternating signal features), or the inherent physics and modulation triggers of power electronics systems could be further leveraged, so as to enhance the model interpretability under spike sparsity.

    \item \textit{Noise resiliency}: The noise resiliency of an SNN is primarily associated with whether the \textit{events} (spikes) are over-captured due to the perturbations with the given thresholds \cite{SpikeTalkNoise}. Representative system performances are tested under different noise levels in Fig.~\ref{fig_noise}, where Additive White Gaussian Noise (AWGN) is applied to the sampled current signal $\mathrm{\mathbf{i}_\mathrm{abc}}$. The results align with the ideal case when the signal-to-noise-ratio (SNR) is high enough to prevent critical spurious spikes, whereas the constructed SNN may fail to converge in excessive noisy cases. On one hand, $\mathrm{\Delta\mathbf{i}_\mathrm{abc}}$ in Fig.~\ref{fig_current} exhibits regular periodic patterns, and encoding with $\mathrm{\Delta\mathbf{i}_\mathrm{abc}^2}$ in (\ref{eq_Iin}) allows the signal to be separated from noise with inferable power spectrum, thereby assisting filter design via signal tapering for best performance. On the other hand, the minimum firing threshold should also be tuned to sufficiently exceed the dominant noise level, so that the spike trains could be effectively triggered by genuine signal fluctuations.

    \begin{figure}[t]
        \centering
        \includegraphics[width=\linewidth]{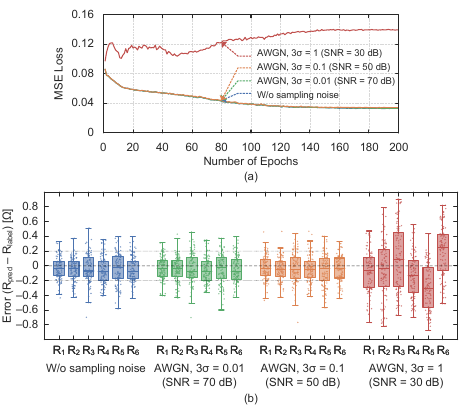}
        \vspace{-12pt}
        \caption{Evaluation of noise resilience: Testing under different sampling noise conditions with 100 random samples in each case, (a) Mean-square-error (MSE) loss in relation to the number of epochs for training, and (2) Distribution of prediction after 200 epochs. The signal-to-noise ratio (SNR) is calculated based on the nominal current rating $\mathrm{I_\mathrm{abc}^* = 10\:A}$.}
        \label{fig_noise}
        \vspace{-9pt}
    \end{figure}

    \item \textit{Device-neuron mapping}: Seen from the neuron model, a more fundamental extension lies in the formalization of device-neuron mapping. Specifically, if $\mathrm{R_{dson}}$ and $\mathrm{C_{ds}}$ are mapped to the membrane resistance and capacitance of a spiking neuron, as both physically share the common $\mathrm{RC}$ units \cite{8930563}, then the neural network can be constructed with reduced number of neurons and trained from few samples for better adaptability, potentially maintaining good prediction performance \cite{SpikeTalk}.

    \item \textit{Event-driven features and online monitoring}: Moreover, the event-driven nature of SNNs enables a further extension toward online monitoring. Spike-Timing-Dependent Plasticity (STDP) enables the online training \cite{TrainingSNN}, allowing the model to adapt to the aging of individual devices during operation with less need for offline training. Also, the event-driven scheme implies the functionality of SNN reflecting the accumulation of critical events, e.g., critical changes in $\mathrm{R_{dson}}$ that accumulate into the alert of state of health or end of lifetime. As computation would be concentrated around those critical events, such monitoring could further enhance the energy efficiency at the edge.
\end{enumerate}

\section{Conclusions}

This article presents a novel neuromorphic-inspired approach for condition monitoring in power electronics. The approach exploits the principle of latency coding of spiking neurons to infer the \textsc{on}-state resistances of power devices through an SNN. The presented results demonstrate adequate prediction performances, underlying constraints and possible improvements, while this approach also implies potential in achieving high energy efficiency during inference and low requirements on data acquisition. Moreover, deeper insights are also provided outlining future extensions upon the intrinsic physics, benefiting the applicability of the methodology.

\section*{Acknowledgement}

The authors acknowledge the support provided by the BMW Group, Munich, Germany, in relation to the project \textit{Reliability and Lifetime Analysis for Electronic Components}.

\bibliography{references.bib}
\bibliographystyle{IEEEtran}

\end{document}